\documentclass[preprint]{aastex631}
\usepackage{algorithm,algpseudocode}

\begin{document}

\title{Modeling of Ionization and Recombination Processes in Plasma with Arbitrary Non-Maxwellian Electron Distributions}

\correspondingauthor{Chengcai Shen}
\email{chengcaishen@cfa.harvard.edu}

\author[0000-0002-9258-4490]{Chengcai Shen}
\affiliation{Center for Astrophysics \textbar\ Harvard \& Smithsonian, 60 Garden Street, Cambridge, MA 02138, USA}

\author[0000-0001-5278-8029]{Xiaocan Li}
\affiliation{Dartmouth College, Hanover, NH 03750, USA}

\author[0000-0002-8747-4772]{Yuan-Kuen Ko}
\affiliation{Space Science Division, Naval Research Laboratory, Washington, DC, USA }

\author[0000-0002-7868-1622]{John C. Raymond}
\affiliation{Center for Astrophysics \textbar\ Harvard \& Smithsonian, 60 Garden Street, Cambridge, MA 02138, USA}

\author[0000-0003-4315-3755]{Fan Guo}
\affiliation{Los Alamos National Laboratory, Los Alamos, NM 87545, USA}

\author[0000-0002-4980-7126]{Vanessa Polito}
\affiliation{Lockheed Martin Solar and Astrophysics Laboratory, Palo Alto, CA 94306, USA}
\affiliation{Department of Physics, Oregon State University, 301 Weniger Hall, Corvallis, OR 97331}

\author{Viviane Pierrard}
\affiliation{Royal Belgian Institute for Space Aeronomy (BIRA-IASB), Space Physics, Solar Terrestrial Center of Excellence (STCE), Brussels, Belgium}
\affiliation{Earth and Life Institute—Climate Sciences (ELI-C), Université Catholique de Louvain, Louvain-la-Neuve, Belgium} 



\begin{abstract}

In astronomical environments, the high-temperature emission of plasma mainly depends on ion charge states, which requires accurate analysis of the ionization and recombination processes.
For various phenomena involving energetic particles, the non-Maxwellian distributions of electrons exhibiting high-energy tails can significantly enhance the ionization process. 
Therefore, accurately computing ionization and recombination rates with non-Maxwellian electron distributions is essential for emission diagnostic analysis.
In this work, we report two methods for fitting various non-Maxwellian distributions by using the Maxwellian decomposition strategy.
For standard $\kappa$ distributions, the calculated ionization and recombination rate coefficients show comparable accuracy to other public packages.
Additionally, our methods support arbitrary electron distributions and can be easily extended to updated atomic databases. 
We apply the above methods to two specific non-Maxwellian distribution scenarios: (I) accelerated electron distributions due to magnetic reconnection revealed in a combined MHD-particle simulation; (II) the high-energy truncated $\kappa$ distribution predicted by the exospheric model of the solar wind.
During the electron acceleration process, we show that ionization rates of high-temperature iron ions increase significantly compared to their initial Maxwellian distribution, while the recombination rates may decrease due to the electron distribution changes in low-energy ranges. 
This can potentially lead to an overestimation of the plasma temperature when analyzing the Fe emission lines under the Maxwellian distribution assumption.
For the truncated $\kappa$ distribution in the solar wind, our results show that the ionization rates are lower than those for the standard $\kappa$ distribution, while the recombination rates remain similar. This leads to an overestimation of plasma temperature when assuming a $\kappa$ distribution.
\end{abstract}



\section{Introduction} \label{sec:intro}

The ionization and recombination processes in astronomical plasma are fundamental to the study of various plasma emissions. In high-temperature optically thin plasma (e.g., the solar corona at $\sim 10^6$ K), the radiation is due to the emission of photons from level transitions in ions of different charge states by electron collisions, resulting in distinct spectral lines. And the intensity of a spectral line is proportional to the ion fraction of the respective ion.
Therefore, an accurate analysis of ionization and recombination processes is essential for understanding plasma properties and serves as an important diagnostic method in various astrophysical and space studies.
The ion charge state in ionization equilibrium is closely correlated to the plasma temperature as the ionization and recombination rates are generally more sensitive to temperature (and less to density). 
Consequently, the populations of ions with different charge states can indicate the plasma temperature, while specific line ratios within single ions are used to estimate the plasma density. 
This is an important diagnostic technique in solar corona observations, where emission lines (and their ratios) provide critical information about the coronal plasma.

Furthermore, the non-equilibrium ionization (NEI) process commonly happens in various environments where the ion charge states provide information about the plasma heating and cooling histories. 
In fact, the high-temperature structure is generally associated with highly dynamic flows in various environments. 
For instance, during the fast magnetic reconnection in solar flares, the reconnection outflows can be as high as $1000$ km/s based on both theoretical predictions and observational evidence.  The temperature of the plasma in reconnection sites often changes on a time scale shorter than the characteristic time scales for ionization and recombination.
Once the plasma flows into the reconnection regions, the cool plasma can be heated rapidly, for instance, as it accesses the Petschek-type slow mode shocks.  
The charge states of the plasma, therefore, correspond to a lower temperature until ionization catches up. 
On the other hand, heated plasma flowing with reconnection outflow in exhausting regions could rapidly cool due to expansion once the heating is not comparable. That will cause the charge states to indicate a higher temperature until recombination catches up, showing over-ionized features \citep{Shen2023ApJ...943..111S}.
In such situations where the plasma is either under- or over-ionized, it is necessary to determine the charge states by solving time-dependent ionization equations with precise ionization and recombination rates.

In general, the ionization and recombination processes are computed with the assumption of Maxwellian electron distributions, and the relevant ionization and recombination rates, as well as ionization states,  can be found in publicly available atomic databases (e.g., CHIANTI \citep{Dere_1997A&AS..125..149D, Dere_2019ApJS..241...22D}, AtomDB \citep{Foster_atomdb_2018}).
These atomic databases contain detailed ionization and recombination data as well as emission tables.
However, the ionization and recombination rate coefficients for various electron velocity distributions other than Maxwellian might be significantly different and require specific treatment as the electrons might be accelerated in different processes. 

In the solar corona, a significant fraction of the electrons can be accelerated into a power-law energy spectrum  \citep[e.g., a spectral index $3 \sim 9$~, in][]{Effenberger2017Hard}. 
Although the dominant acceleration mechanisms are not yet fully understood—whether they include direct electric field, turbulence, or contracting plasmoid \citep{Guo2024SSRv..220...43G}-the particle distributions exhibiting high-energy tails can be predicted in different particle acceleration processes.
In macro-scale systems such as in magnetic reconnection sites, the acceleration of particles has been widely investigated based on analytical modeling and MHD simulations \citep[e.g.,][]{Zank_2015ApJ...814..137Z, Montag_2017PhPl...24f2906M, Li_2018ApJ...866....4L}. 
In solar observations, non-thermal signals with high-energy power-law tails are also routinely observed in hard X-ray (HXR) spectra of above-the-loop sources during flares \citep[e.g.,][]{krucker08}. 
In some cases, these spectra have been fitted with a kappa ($\kappa$) distribution \citep[e.g.,][]{oka13}. 
In this way, $\kappa$ functions provide a convenient description of non-Maxwellian distributions using a free parameter $\kappa$ ranging from $\sim$~1.5 to infinite,
and arise naturally from Tsallis statistical mechanics \cite[e.g.][]{livadiotis13}. 
In practice, observational signatures of $\kappa$ distributions for the electrons have been found using intensity ratios of optically thin lines observed by EUV spectrometers, such as Hinode/EIS \citep[e.g.,][]{dzifcakova10,dzifcakova12} and SDO/EVE \citep{dzifcakova18}.
These lines are primarily influenced by ionization, recombination, and excitation processes that occur through electron–ion collisions, which are highly dependent on the electron velocity distribution. Thus, the presence of non-Maxwellian distributions can significantly affect the formation of lines from highly ionized Fe atoms observed by EIS.
More recently, \cite{jeffrey16,jeffrey17} and \cite{polito18} have also shown that in some flare observations, the profiles of EIS high-temperature lines strongly depart from a Gaussian shape and can be best fitted with a $\kappa$ distribution. 
In addition, other forms of non-Maxwellian particle distributions have been frequently reported in the solar wind and Earth's magnetosphere \citep[e.g.,][]{Maksimovic2005JGRA..110.9104M}. The electron distribution function has been widely described by the combination of the Maxwellian core at low energies and high energy tails representing the suprathermal power law.

A set of extended packages based on popular atomic databases has been developed to calculate the ionization and recombination rates using Kappa distributions. For instance, the KAPPA package facilitates the calculation of optically thin emission line and continuum spectra for solar and stellar coronae with kappa distributions, utilizing data from CHIANTI \citep{Dzifcakova2015ApJS..217...14D,Dzifcakova2021ApJS..257...62D}. Similarly, AtomDB \citep{Foster_atomdb_2018} provides associated packages to calculate ionization and recombination rates using the Maxwellian decomposition method and associated tables reported by \cite{Hahn2015ApJ...809..178H}.
However, these databases and packages are either optimized for specific kappa values, such as 1.7 to 100 in AtomDB, or require a particular version of the CHIANTI database, thereby limiting their applicability in various modeling scenarios. Moreover, the current packages mainly focus on the standard kappa distribution, not covering cases where other non-Maxwellian distributions might be significant.
Therefore, it is useful to develop a more flexible method to accurately calculate ionization and recombination rates in various non-Maxwellian distributions beyond the standard kappa distribution.

In this paper, we present two approaches to efficiently calculate ionization and recombination rates using the Maxwellian decomposition method. To improve the accuracy of computations, we employ modern machine-learning libraries in our calculations.
In Section 2, we provide a brief review of the Maxwellian decomposition method and describe two fitting methods employed in our calculations. In Section 3, we discuss the calculation results for standard kappa distributions and provide two specific examples of other non-Maxwellian distributions expected in magnetic reconnection sites and solar winds. Finally, in Section 4, we summarize our results and discuss potential applications of this method.

\section{Numerical Method: Maxwellian Decomposition} \label{sec:method}
\subsection{Kappa Distribution}
In this work, we apply the Maxwellian decomposition method to fit any non-thermal distribution as a sum of multiple-Maxwellian components \citep[e.g.,][]{Ko1996GeoRL..23.2785K, Anderson1996, Kaastra2009A&A...503..373K,  Hahn2015ApJ...809..178H}.
In this way, an arbitrary elected distribution function, $f(E)$, can be represented as 
\begin{equation}
    f(E) = \sum c_i f_M(E, T_i),
\end{equation}
where $f(E)$ is the fraction of electrons with energy between $E$ and $E + dE$,  $c_i$ is a normalizing coefficient of the $i-$th Maxwellian component at temperature $T_i$. Here, the Maxwellian distribution is given by
\begin{equation}
    f_M(E, T_i) = \frac{2}{\sqrt{\pi}}\left(\frac{1}{k_BT_i}\right)^{3/2}\sqrt{E}\exp\left(\frac{-E}{k_BT_i}\right).
\end{equation}
The ionization and recombination rate coefficients for the distribution function $f(E)$ can be obtained by summing the Maxwellian rate coefficient at the temperature $T_i$.

We first focus on the most commonly used non-Maxwellian distribution of electrons, the kappa distribution.
The widely used kappa definition of the velocity distribution function (VDF), Olbertian Kappa distribution \citep{Olbert1968ASSL...10..641O}, 
is first introduced to describe the electron VDF in studies of magnetospheric electron spectral measurements. Further investigations with various kappa forms, such as modified kappa distributions, have been reviewed by \cite{Fichtner2021} in detail. 
In this work, we apply the following form of Kappa distribution \citep[e.g.,][]{Owocki1983ApJ...270..758O,Dzifcakova2015ApJS..217...14D, Hahn2015ApJ...809..178H}:

\begin{equation}
    f_{\kappa}(E) = A_{\kappa}\frac{2}{\sqrt{\pi}}\left(\frac{1}{k_BT}\right)^{3/2}\sqrt{E}\left(1+\frac{E}{(\kappa-1.5)k_{B}T}\right)^{-\kappa-1},
\end{equation}
where $k_B$ is the Boltzmann constant, and $T$ is the electron temperature. 
Here $A_{\kappa}$ is a normalization constant defined by:
\begin{equation}
    A_{\kappa} = \frac{\Gamma(\kappa+1)}{\Gamma(\kappa-1/2)(\kappa-1.5)^{3/2}}.
\end{equation}
where $\Gamma$ is the Gamma function. 
Alternatively, we can represent $E$ in units of $k_BT$ as $E^{'}=E/(k_BT)$. Thus, the distribution function can be conveniently described by $f_{\kappa}^{'}(E^{'})=A_{\kappa}\frac{2}{\sqrt{\pi}}\sqrt{E^{'}}\left(1+\frac{E^{'}}{(\kappa-1.5)}\right)^{-\kappa-1}$, and $f_{\kappa}(E) = \frac{f_{\kappa}^{'}(E^{'})}{k_BT}$. 
In the following section of this paper, the task would be finding a set of suitable Maxwellian components of $f_{\kappa}^{'}(E^{'})$. Thus, these Maxwellian component coefficients ($c_i$) are the same for the standard $\kappa$ distribution at any temperature.

\begin{figure}[ht!]
\centering
\includegraphics[width=0.6\textwidth]{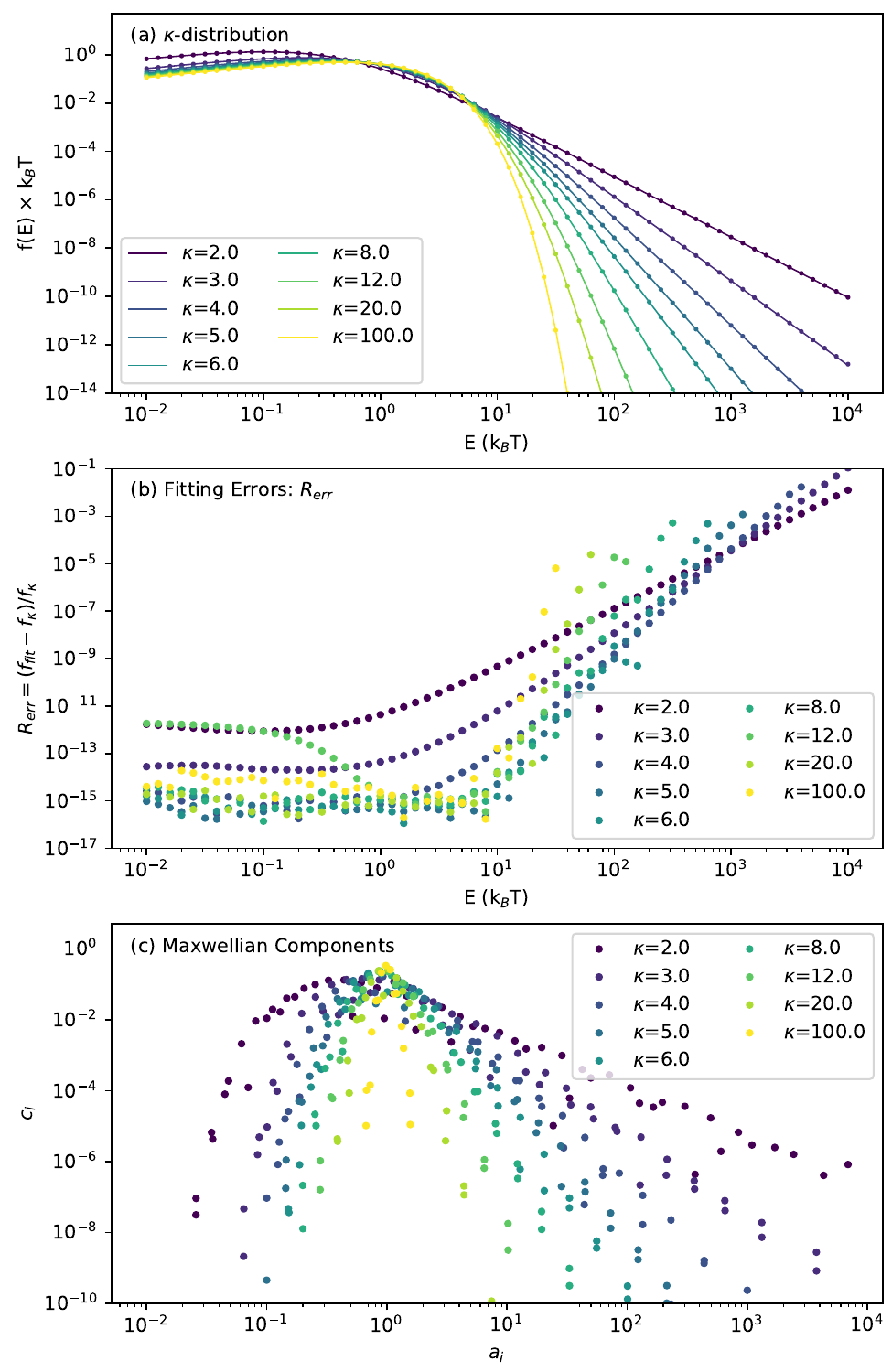}
\caption{The Maxwellian decomposition fitting results for the chosen $\kappa$-distribution profiles. In panel (a), the solid lines are kappa distribution $f(E)$ and the dots show the results by summing multiple Maxwellian components. The $E$ is in units of $k_BT$ (eV) in this plot;  (b) The fitting errors defined by $|f_{fitting} - f_{\kappa}|/f_{\kappa}$ for each kappa case; (c) The Maxwellian components coefficients: $c_i$ and $a_i$. The summing distribution profile is obtained by $f(E)_{fitting} = \sum c_if(a_i T)_{Maxwell}$.
\label{fig:linearfit}}
\end{figure}

\subsection{Classic Iterative Method}
To find the proper Maxwellian components, we employ the classic iterative method to solve the following linear least-squares problem:
\begin{equation}
    \text{minimize }  \|f(E) - \sum c_i f_M(E, T_i)\|^2.
\end{equation} 
By initializing a set of temperature grids ($T_i$) and associated Maxwellian profiles with $T_i$, a fitting process seeks coefficients $c_i$ to minimize the residual sum of squares between the target non-Maxwellian distribution $f(E)$ and the sum of multiple Maxwellian components ($\sum c_i f_M(E, T_i)$).
We utilize LSQR method \citep{Paige1982ACMTM...8...43P} to perform the fitting, which has been packaged in the popular machine learning library, \textit{Scikit-learn.LinearRegression}. 
LSQR is a conjugate-gradient type scheme for solving sparse linear equations and sparse least-squares problems. It is based on the Golub-Kahan bidiagonalization process, and analytically equivalent to the standard method of conjugate gradients but has better numerical properties. 
In particular, LSQR avoids explicit matrix factorization. Thus, it is well-suited for ill-conditioned problems where numerical errors or noise may have significant impacts on the fitting process. 
The LSQR solver can fit a distribution profile with extremely high accuracy (e.g., to the precision of a double floating type).
The \textit{Scikit-learn.LinearRegression} also supports solving the KKT (Karush-Kuhn-Tucker) conditions for the non-negative least squares problem to explicitly enforce the non-negativity constraints on $c_i$.
As LSQR performs well for large-scale sparse least-squares problems, we are able to set up a large number of temperature grids in the fitting for the standard Kappa distributions (e.g., up to $\sim 10^4$\ $T_i$ nodes), which is equivalent to a high resolution of $\Delta LogT \sim 0.0008$\ K in following fitting processes.

\begin{figure}[ht!]
\centering
\includegraphics[width=0.6\textwidth]{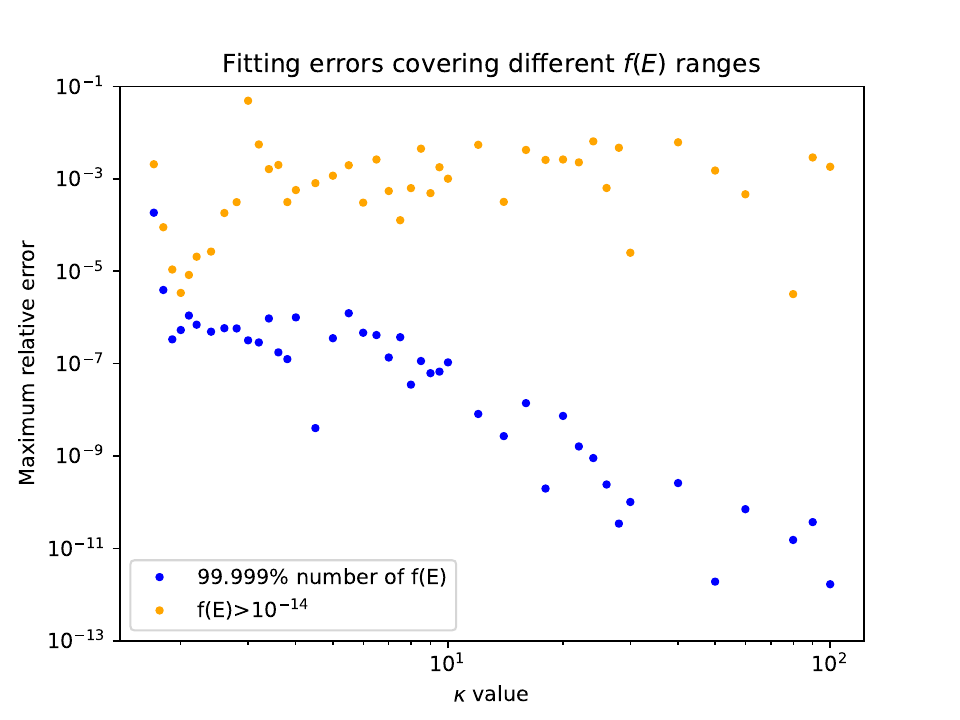}
\caption{
The maximum fitting errors covering different electron distribution profiles $f(E)$ versus $\kappa$-values. The blue dots are results including $f(E)$ with at least 99.999\% number of electrons at the lower energy side, and the orange dots count all ($E$) points with $f(E) > 10^{-14}$.
\label{fig:linearfit_error}}
\end{figure}

Figure \ref{fig:linearfit} shows the fitting result for the standard Kappa distribution with $\kappa$-values ranging from 2 to 100. It is clear that the multiple-Maxwellian components approximation (dot curves) match the $\kappa$ profiles (solid lines) with very high accuracy. 
Here, the relative error between the ${f_{fitting}}$ and kappa function is as low as the order of $10^{-15}$ for the dominant ranges of electron population ($f(E)$) as shown in Figure \ref{fig:linearfit}(b). 
The corresponding coefficients, $c_i$ and $a_i$, are displayed in Figure \ref{fig:linearfit}(c). As expected, the small kappa cases (e.g., $\kappa=2$) require Maxwellian components at finer temperature grids ($a_i$), which range over 4 orders of magnitude. While the large kappa case (e.g., $\kappa=100$) can generally be well fitted with fewer Maxwellian components.
The fitting errors for all kappa values from $\kappa=1.7$ to $\kappa=100$ are also plotted in Figure \ref{fig:linearfit_error}. 
Here, we estimate the maximum relative errors for the selected $f(E)$ in two conditions: (I) for the dominant part of $f(E)$ covering at least 99.999\% of the electron numbers at different energies starting from the low energy end; (II) for all electron energy ($E$) points with $f(E) > 10^{-14}$, approaching the threshold of double-precision float type in numerical solvers.
For condition (I), all relative errors are less than $10^{-3}$ and continuously decrease to even $\sim 10^{-11}$ at large kappa cases.
For the condition (II), the errors also achieve a level of approximately $10^{-3} \sim 10^{-2}$, even with the condition of $f(E) > 10^{-14}$ as shown by the orange curve in Figure \ref{fig:linearfit_error}.

\subsection{Linear Regression Enhanced by ANNs}
The second method we utilized is a specified linear regression model optimized by the general artificial neural network (ANN) algorithm.
The purpose of this model is to offer more flexibility during the fitting process compared with direct linear algebra methods.
In practice, the above classic iterative method requires pre-defined temperature grids (e.g., $a_i$), which are fixed during the fitting process. 
Thus, to achieve successful fitting for any electron distribution profiles that have not been well studied, a finer temperature grid is required.
However, a large number of temperature grids might cause huge computational costs even for the highly efficient LSQR scheme and sometimes cause the system to crash due to limitations in computer memory (see details in Appendix A and associated figures).
In addition, the choice of parameter range, such as the minimum and maximum of $a_i$, also depends on experience.
Therefore, a specified linear regression model with adaptively adjusted $a_i$ and $c_i$ during the fitting process offers an alternative approach.

\begin{figure}[ht!]
\centering
\includegraphics[width=0.99\textwidth]{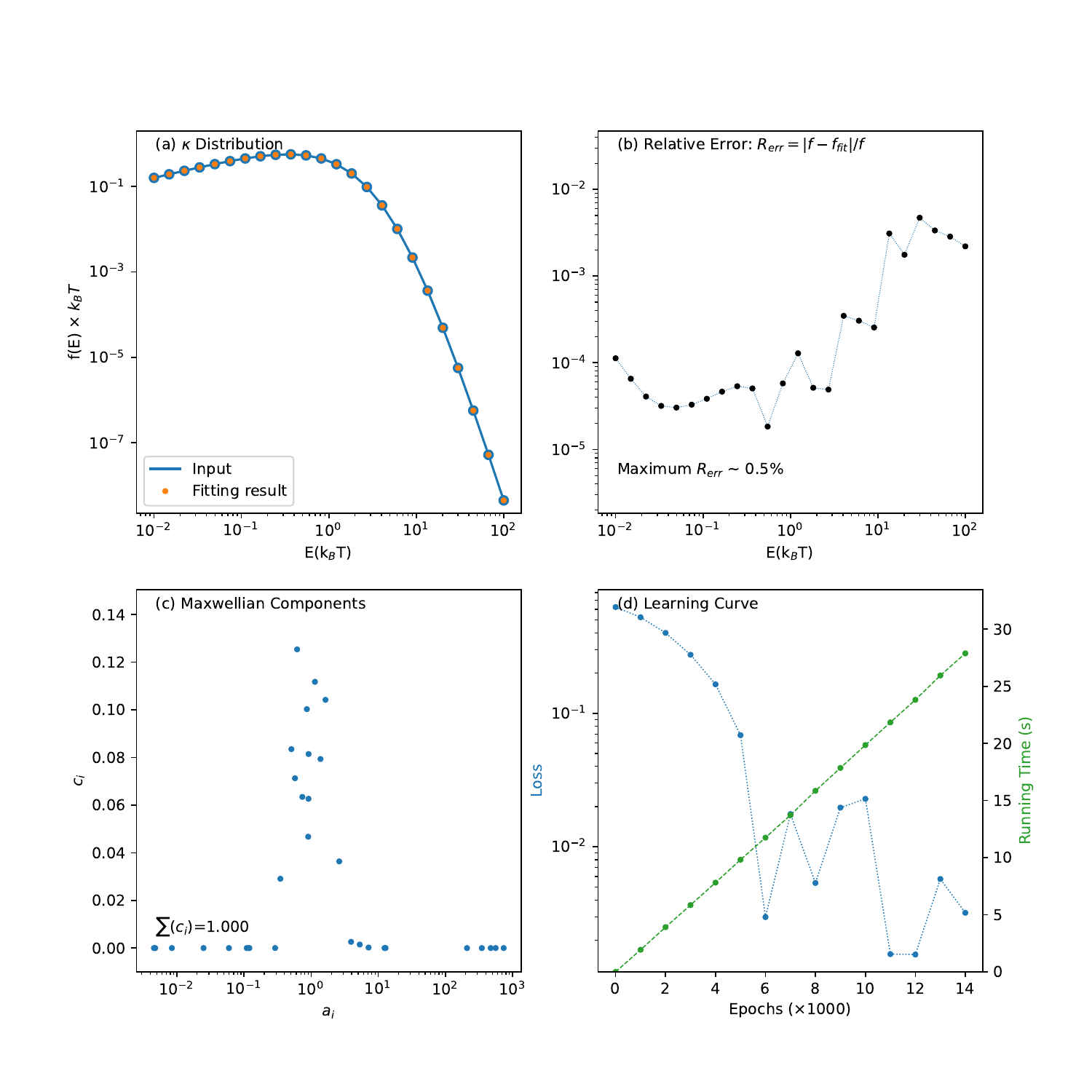}
\caption{The Maxwellian decomposition fitting by employing the neural network model for a typical $\kappa$-distribution. In panel (a), the solid line is $\kappa=6$ distribution $f(E)$ and the dots show the results by summing multiple Maxwellian components from the training results. (b) The relative error between the training results and input defined by $|f(E)_{fitting} - f(E)|/f(E)$; (c) The trainable parameters (or Maxwellian decomposition coefficients in particular in this work): $c_i$ and $a_i$; 
(d) The learning curve shows a mean loss in the training prediction and the target profile $f(E)$. The green dashed line shows the running time.
\label{fig:deepfit_k6}}
\end{figure}

Modern ANN libraries offer an efficient way to build a specialized linear regression model.
A typical three-layer artificial neural network generally consists of an input layer, a hidden layer, and an output layer. 
The input layer connects to the output layer through a hidden layer with two linear combinations, incorporating a non-linear function to add the complexity of the network for realistic applications.
Inside this network, forward functions define how the input data is passed through the network and then pass the result to the output layer. 
In this work, the main goal is to obtain the weight of each linear component, which is represented by $c_i$ as mentioned above.
Therefore, this linear regression model is a particular single-layer neural network in which the input layer is directly connected to the output layer through the forward functions.
In this way, we can build a linear regression model by utilizing well-developed Python libraries such as \textit{PyTorch}, only using specific components from general network models.
The learnable parameter interactions still follow the rules of general neural networks, allowing for the usage of all optimization algorithms during the training process.

We then employ this model on the application to fit the standard Kappa distribution as shown in 
Figure \ref{fig:deepfit_k6}.
The fitting result matches the $\kappa=6$ profile very well, as shown in Figure \ref{fig:deepfit_k6}(a), and the corresponding Maxwellian components are plotted in panel (c).
One of the advantages of this linear regression model is that the normalization condition ($\sum (c_i) = 1$) has been strictly satisfied as it has been examined at each iteration step during the fitting process.
The accuracy of the fitting can be explicitly specified in this linear regression model, which is another obvious advantage.
As shown in Figure \ref{fig:deepfit_k6}(b), one can easily set up the error levels (e.g., 0.01 in this run), and the maximum relative difference between the input and fitted profile is then restricted to $< 1$\%. 
The corresponding computation time and learning curve are shown in Figure \ref{fig:deepfit_k6}(d).
It can be seen that the loss (e.g., the sum of squared errors) decreases quickly for about 2 orders within the initial $6000$ epochs utilizing the default Adam optimizer.
It is worth noting that the choice of the optimizer and learning rate might significantly affect the training speed.
\footnote{The run with $\sim$ 15000 epochs completes in about 30 seconds using a laptop-grade CPU of an Apple’s M3 chip. It should be noted that the running times also depend on the number of samplings to fit. In the above test, a total of 24 sampling points along the $f(E)$ profile were fitted utilizing 32 Maxwellian components.}
By setting the proper error thresholds, this approach also helps to avoid overfitting on the sampling points with minor values close to the numerical precision.
In addition, it benefits from flexible parameter adjustment which is useful in particular fitting applications. For example, it is easy to specify the allowed ranges for $c_i$ during the fitting process.
This makes the linear regression model with ANNs a valuable alternative tool for complex fitting tasks. 

\section{Results} \label{sec:results}
\subsection{Ionization and Recombination Rates in Standard Kappa Distribution}

\begin{figure}[ht!]
\centering
\includegraphics[width=0.8\textwidth]{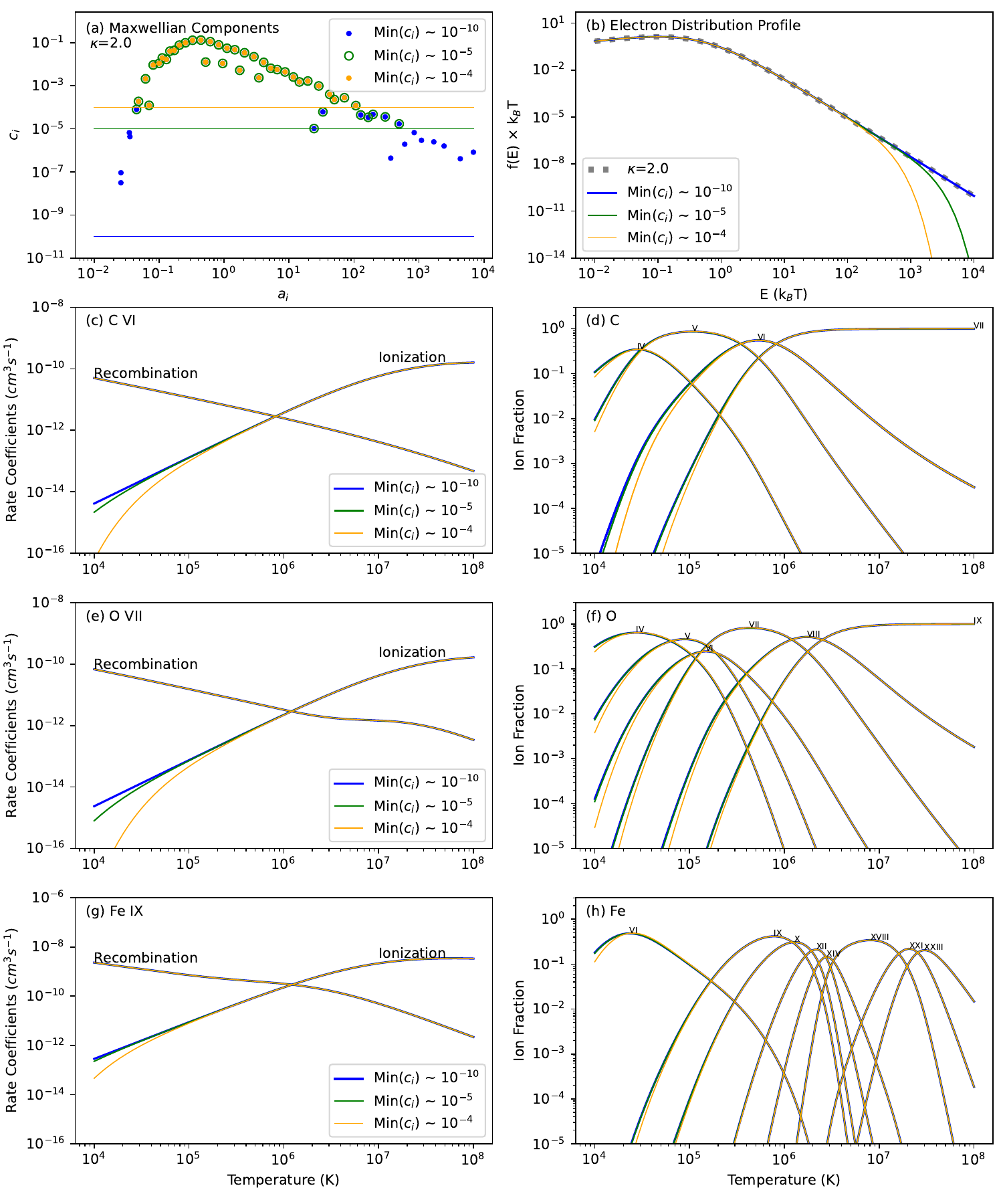}
\caption{The ionization and recombination rates calculated by using different Maxwellian decomposition coefficients for the $\kappa = 2$ case. (a) The chosen Maxwellian decomposition coefficients with different accuracy levels: low to $10^{-10}, 10^{-5}$, and $10^{-4}$ as shown by blue, green, and orange lines, respectively; (b) The $\kappa = 2$ and fitted electron distribution profiles; The following panels (c-d), (e-f), and (g-h) display the corresponding ionization and recombination rate coefficients for ions C VI, O VII, and Fe IX, and their ion fractions in ionization equilibrium assumptions.
\label{fig:rate_k2}}
\end{figure}

After obtaining the Maxwellian decomposition coefficients ($a_i, c_i$), calculating the ionization and recombination rate coefficients is straightforward. This involves adding up all corresponding linear combinations of Maxwellian rate coefficients as:
\begin{equation}
    \alpha_{\kappa} = \sum{c_i \alpha(T_i)},
\label{eq:rate}
\end{equation}
where $\alpha(T_i)$ is the rate coefficient at any temperature $T_i \equiv a_iT$ depending on Maxwellian decomposition coefficients ($a_i, c_i$).
Figure \ref{fig:rate_k2} shows how the distribution of ionization and recombination rates vary with temperature for the $\kappa=2$ distribution. It is noticed that the Maxwellian decomposition method normally contains a large range of coefficients $a_i$ over several orders, which may require the Maxwellian rates to be calculated at very large temperature $T_i$ ranges as well. 
In cases where the temperature ($T_i$) falls outside the current tabulated atomic data, such as CHIANTI, we will perform extrapolation (and interpolation) because the asymptotic behaviour of Maxwellian ionization rates is well known when ignoring the density effects.
It is also straightforward to similarly calculate emissivities of spectral lines of a specified ion over given ranges of temperature and density by taking a weighted sum of the emissivities in Maxwellian distributions, following the methods reported in previous studies \citep[e.g.,][]{shen2013ApJ...773..110S, Hahn2015ApJ...809..178H, shen2017ApJ...850...26S}.

It is important to examine how the accuracy of Maxwellian decomposition affects derived ionization and recombination rates.
The uncertainties in calculations of ionization and recombination rates were discussed in detail by \cite{Hahn2015ApJ...809..178H}. 
Here, we emphasize that the fitting errors at the high-energy end of electron distribution profiles should be carefully considered.
As shown in Figure \ref{fig:rate_k2}(a), we artificially set the accuracy threshold of $c_i$ to $10^{-10}, 10^{-5}$, and $10^{-4}$ and only choose the dominant part of these Maxwellian components in the following tests.
The corresponding fitted electron distribution profiles are then calculated by using the chosen $c_i$ patches above and shown in Figure \ref{fig:rate_k2}(b).
It is clear that the case with $Min(c_i) \sim 10^{-10}$ returns fitting results that align with the $f(E)$ of $\kappa=2$.   
In comparison, the case with $Min(c_i) \sim 10^{-5}, 10^{-4}$ missed a set of electrons at the high energy tails around $E > \sim 10^3 (k_BT)$, as shown by the green and orange lines.

\begin{figure}[ht!]
\centering
\includegraphics[width=0.8\textwidth]{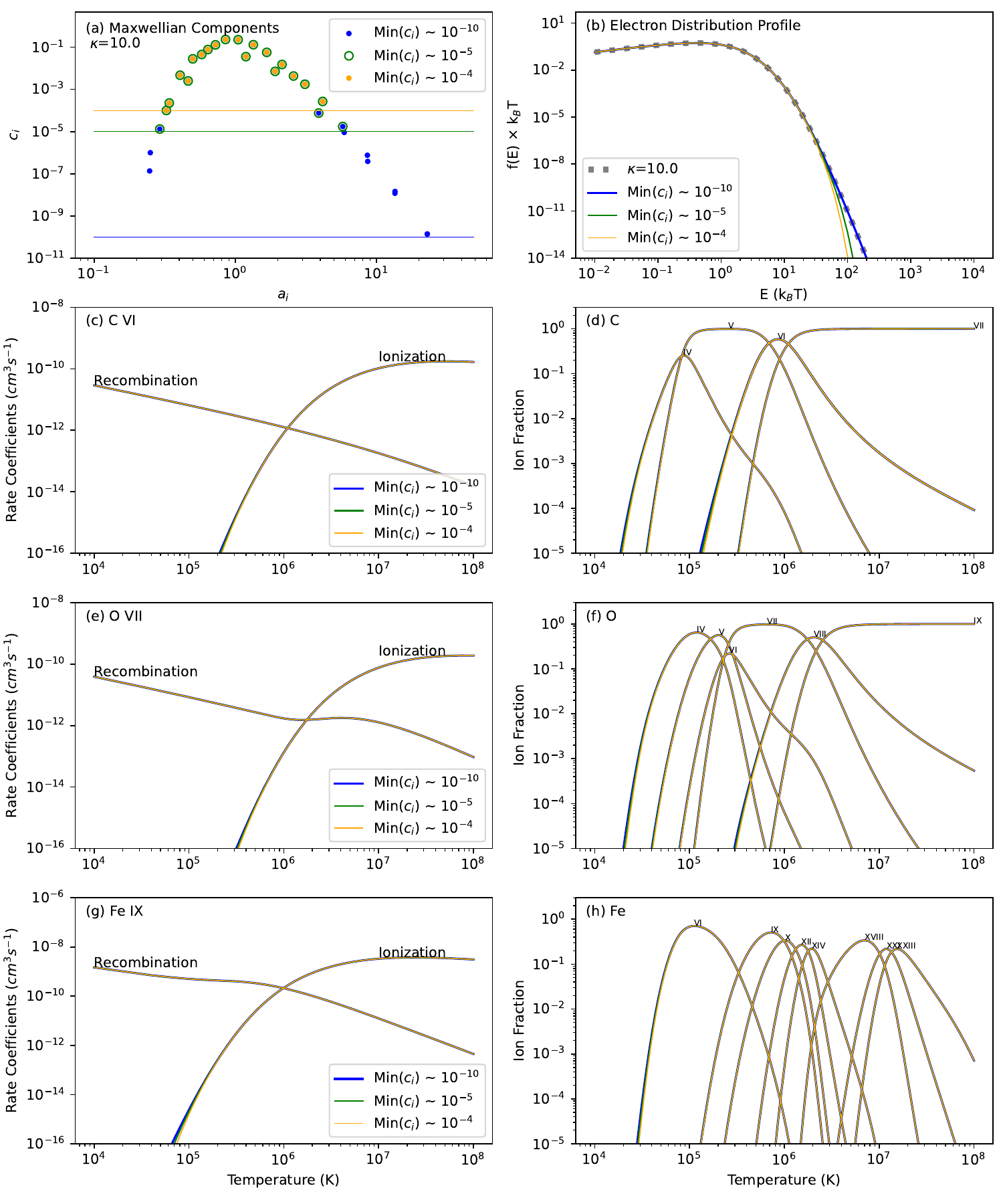}
\caption{Same as in Figure \ref{fig:rate_k2}, but for $\kappa=10$.
\label{fig:rate_k10}}
\end{figure}

The corresponding ionization and recombination rates for C VI, O VII, and Fe IX from the above-chosen Maxwellian components are shown in Figure \ref{fig:rate_k2}(c,e,g). 
It is clear that these components with different accuracy levels ($Min(c_i)$) have significant impacts on ionization rates, especially at low temperatures (e.g., $T < 10^5$ K).
The inaccuracy in fitting on the high-energy end of the electron distribution profiles could cause under-estimation of ionization rate coefficients, as shown by the green and orange lines.
The most significant impacts appear on the ions corresponding to lower temperatures, such as C VI and O VII, while the high-temperature Fe IX ions are less affected.
The impact of high-energy tails on ionization rates can be understood by examining the ionization potentials $E_i$, which are 490/739/234 eV for C VI/O VII/Fe IX ions, respectively. The most important aspect is to review how many electrons with energies above the ionization potential or the position of $E_i$ on the distribution function (e.g., along the horizontal axis in Figure \ref{fig:rate_k2}b).
For example, C VI and O VII ions are more easily affected by the high-energy tail (especially in lower kappa cases) compared to Fe IX, especially at low T.
At higher electron temperatures, $E_i$ is close to the `core' of the Kappa distribution function, resulting in less effect of the high temperature tail.
In comparison, recombination rates are not as sensitive to those Maxwellian components, and we can see that recombination rate coefficients have nearly identical profiles over the three cases.

The right panels in Figure \ref{fig:rate_k2} show ion fractions in ionization equilibrium, calculated by using the above selected Maxwellian components. 
For $\kappa = 2$, the calculations using Maxwellian components with $Min(c_i) \sim 10^{-5}$ provide reliable ion fractions, which closely match the results obtained by using more accurate fitting of $Min(c_i) \sim 10^{-10}$. However, when using the Maxwellian components with $Min(c_i) \sim 10^{-4}$, noticeable deviations still exist for some ions, such as C VI and O VI, as shown by oranges lines in Figure \ref{fig:rate_k2}(d) and (f).
A similar analysis for the $\kappa=10$ distribution is shown in Figure \ref{fig:rate_k10}. In this case, the contribution of high energy tail to the overall electron distribution is less compared to the $\kappa = 2$ case, and the impact of inaccuracies of Maxwellian components is minor and can be ignored when examining the ion fractions. 
In summary, to achieve higher accuracy in the ionization and recombination rate calculations, it is important to include even very small Maxwellian components. 
The above analysis suggests that the accuracy of $c_i \sim 10^{-5}$ is a proper criterion for most of the standard $\kappa$ distributions, even for the extreme distribution of $\kappa = 2$. However, it is still worth improving the accuracy of the Maxwellian decomposition process to cover high-energy tails more effectively in high-accuracy simulations.

\begin{figure}[ht!]
\centering
\includegraphics[width=0.8\textwidth]{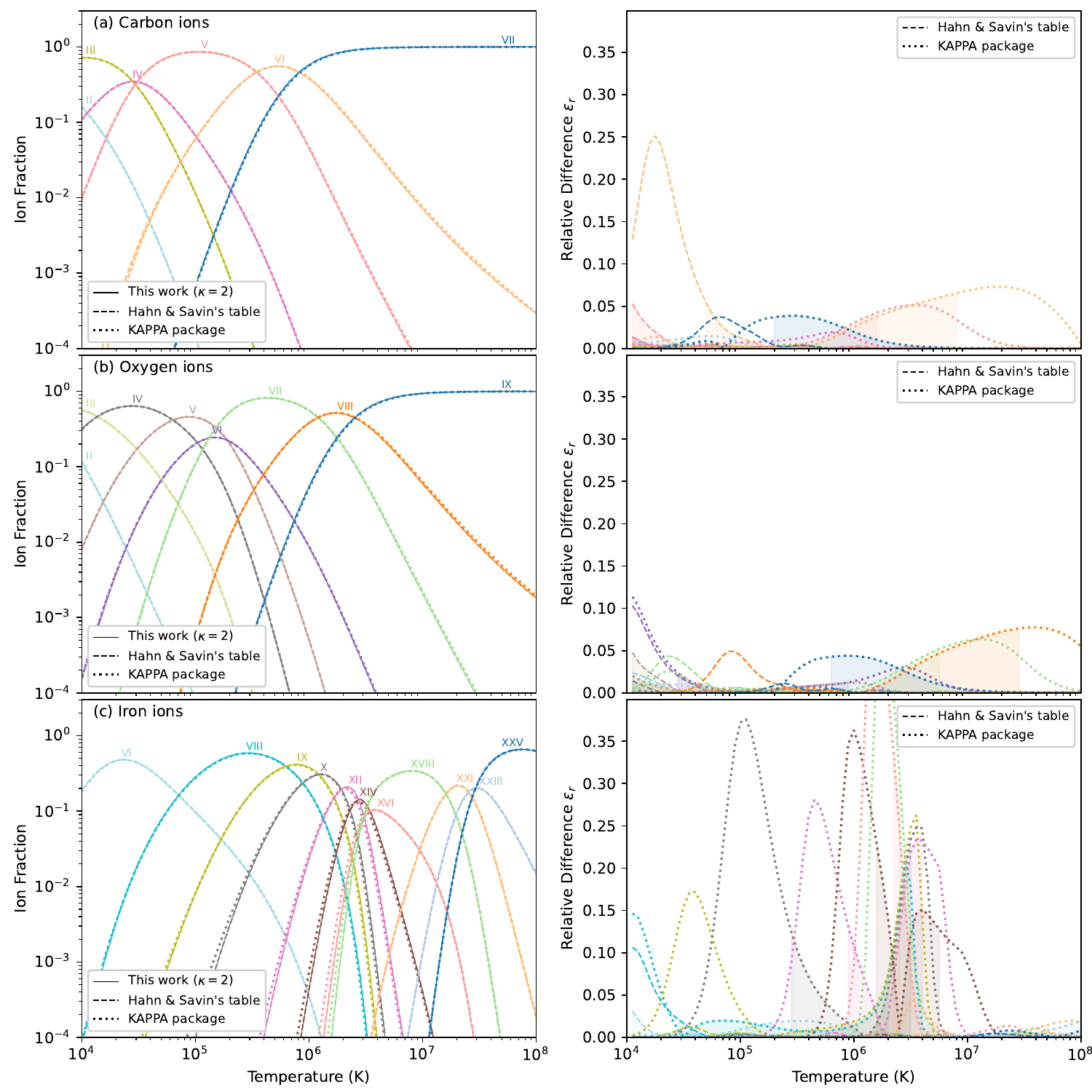}
\caption{Ion population of carbon, oxygen, and iron in ionization equilibrium states for the $\kappa = 2$ case. The left panels show the ion fraction and the right panels are relative differences between our results (solid lines) and others using the table in \cite{Hahn2015ApJ...809..178H} (dashed lines) and the KAPPA package \citep{Dzifcakova2021ApJS..257...62D} (dotted lines). 
The different line colors denote the different ionic charge states. 
The two groups of relative differences are computed by using the formula: $\epsilon_r = |f_{this work}-f_{others}|/Max(f_{this work}, f_{others}+10^{-4})$, respectively.
The shadow regions in the right panels denote the temperature ranges where the ion fraction is substantial as $f > 10^{-2}$.}
\label{fig:ei_fraction_comp}
\end{figure}

The other charge states of carbon, oxygen, and iron obtained by using the above ionization and recombination rates are shown in Figure \ref{fig:ei_fraction_comp}, and the results are compared with others obtained by using the public data, including the tables of \cite{Hahn2015ApJ...809..178H} and KAPPA packages \cite{Dzifcakova2021ApJS..257...62D}.  
The left panels of Figure \ref{fig:ei_fraction_comp} show the ion fractions in ionization equilibrium for the $\kappa=2$ case. Here, solid lines are calculated by using our Maxwellian decomposition results, and dashed and dotted lines are from the table of \cite{Hahn2015ApJ...809..178H}'s paper and KAPPA package with CHIANTI version v10.1, respectively.
For those carbon, oxygen and iron ions with substantial populations, our results well match the results from both \cite{Hahn2015ApJ...809..178H}'s table and the KAPPA package. The peaks of Fe ion distributions match each other as well.
The relative differences between our results and the other two approaches are displayed in the right panels in Figure \ref{fig:ei_fraction_comp}. It is clear that for most carbon and oxygen ions, the relative differences are at the level of a few percent. 
However, the relative differences between our calculations and those of the KAPPA package are noticeable for some Fe ions, particularly for ions from Fe IX to Fe XVIII. Nonetheless, the manifest differences only appear where the ion populations are very low, while the relative differences remain minor for dominant ions (e.g., $f > 10^{-2}$), as indicated by shadow regions in Figure \ref{fig:ei_fraction_comp}. Thus, the differences between our results and KAPPA package do not significantly affect emission diagnostics in practical applications.

\subsection{Ionization and Recombination Rates in Magnetic Reconnection Simulations}
\begin{figure}[ht!]
\centering
\includegraphics[width=1.0\textwidth]{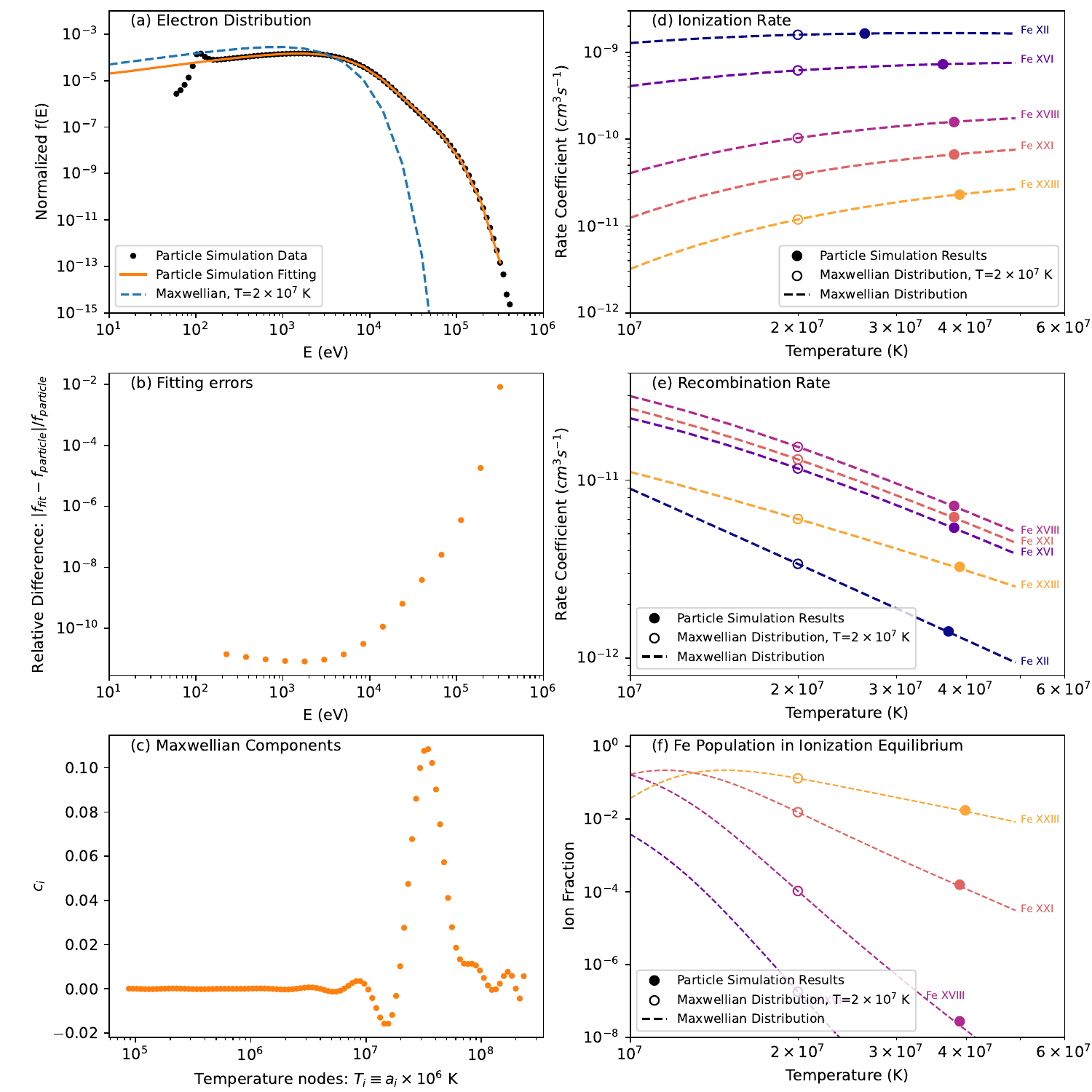}
\caption{The Maxwellian decomposition fitting results for an electron distribution from particle simulations, and the corresponding ionization and recombination rates. 
(a) The normalized electron distribution profiles;
(b) The relative differences between the fitting results and the particle simulation data;
(c) The corresponding Maxwellian decomposition coefficients ($c_i$);
(d-e) Ionization and recombination rate coefficients. Here, the filled circles indicate results from particle simulation data, open circles show the Maxwellian distribution assumption at $T=2 \times 10^7$\ K, and dashed lines are for Maxwellian distribution at different temperatures;
(f) Ion populations (Fe XVIII, XXI, XXIII) in ionization equilibrium states.
\label{fig:simulation_li}}
\end{figure}

In this section, we apply the Maxwellian decomposition method to analyze ionization and recombination in arbitrary electron distributions due to the electron acceleration in magnetic reconnection regions in solar eruptions.
It has been found that the electron distribution in various structures during solar eruptions can be different from a standard $\kappa$ distribution due to rapid dynamic processes, in which the higher energy electrons require different times to gain energy and generally evolve with time \citep[e.g.,][]{Li_2022ApJ...932...92L}. 
For instance, in numerical models, \cite{Li_2022ApJ...932...92L} reported that the electrons can be accelerated to hundreds of keV and the electron distribution approaches to the broken power-law spectral shape. 
Their simulation also revealed that the slope of the high energy tail on the electron distribution varies at different phases during the system's evolution. 
Therefore, it is important to investigate how the ionization and recombination rates are affected by these various high-energy electron tails.

In this work, we obtain non-Maxwellian electron distribution samples by using combined magnetohydrodynamic (MHD) and particle simulations. 
In the MHD simulation, we apply the same parameter as described in \cite{Shen2023ApJ...943..111S} (Case D) to reveal the standard flare geometry in 2D, where a vertical reconnection current sheet naturally develops due to the fast magnetic reconnection associated with the formation of a closed magnetic loop system underneath.
We then solve Parker's energetic particle transport equation to evolve a nonthermal electron distribution based on the temporal-evolutional background magnetic and velocity fields derived from the MHD simulation.
The algorithm is the same as in~\citet{Li_2018ApJ...866....4L,Li_2022ApJ...932...92L}, where the spatial diffusion coefficient is evaluated based on turbulence property assumptions, including the turbulence correlation length ($10^3$\ km) and the normalized turbulence amplitude of 0.2.
The electrons initially injected into the reconnection current sheet regions have a Maxwellian distribution assuming a core temperature of $\sim 1.7$ keV ($\sim 2 \times 10^7$\ K), as shown by the cyan dashed line in Figure \ref{fig:simulation_li}(a). 
As the magnetic reconnection takes place and the primary reconnection X-point appears gradually, super-Alfvenic reconnection outflows are driven along the reconnection current sheet. 
Due to the flow compression in the reconnection current sheet region, 
the electrons are accelerated up to a few hundred keV and develop a nonthermal power-law tail.
We count all accelerated electrons in a nearby area around the reconnection current sheet at a randomly selected time and display the overall electron distribution profile in Figure~\ref{fig:simulation_li} (a), represented by the black dotted curve.
Compared with the initial electron distribution, the high-energy part of the electrons is significantly increased. 
It is clear that the electron distribution at high energy cannot be simplified to a single power-law or simple $\kappa$ function 
as the profile slope is much softer at the high energy end compared with the middle energy range ($E \sim$10-100 keV), which is commonly found in particle simulations. In general, the energy distribution tends to have a power-law (or double power-law) energy range and then appears clear energy cutoff (or exponential rollover) affected by the simulation scales, magnetic guide fields, or escape effects \citep{Guo2024SSRv..220...43G}.
At the low energy end, we made a Maxwellian approximation to the simulation data to reduce the impact of numerical noise ($E \leq 0.1$ keV), as shown by the blue dotted line.

We perform Maxwellian decomposition fitting on the above electron distribution profile by utilizing the classic iterative method as described in Section 2.2.
The fitting results, as shown by the orange solid line in Figure \ref{fig:simulation_li}\ (a), well match the particle simulation data with minor relative errors (Figure \ref{fig:simulation_li}b). 
For comparison, we also show the initial Maxwellian distribution at $T = 2 \times 10^7$\ k as the dashed line, in Figure \ref{fig:simulation_li}\ (a). 
As shown in Figure \ref{fig:simulation_li}\ (c), the dominant Maxwellian components ($c_i$) for the particle simulation data clearly shift to the high-temperature side compared to its initial temperature at $2 \times 10^7$\ K.
Because the simulated profile is not as smooth as the ideal $\kappa$ distribution, the values of $c_i$ from this fitting show fluctuations around zero. 

Figure \ref{fig:simulation_li}(d,e) shows the corresponding ionization and recombination rate coefficients for the chosen Fe ions.
The overall feature is that the ionization rate coefficients for the accelerated electrons (filled circles) are significantly higher than the initial Maxwellian distribution at $T=2 \times 10^7$\ K (open circles).
This is because the ionization process is particularly sensitive to the high-energy tail distribution. In the simulation case, there are more high-energy electrons in the tail, which leads to a higher ionization rate coefficient.
For the chosen high-temperature Fe ions, the dominant recombination process is radiative recombination, which is affected by the number density at the lower energy end. 
As particles are accelerated to higher energy ranges, a decrease in number density at lower energies are clear, leading to a corresponding reduction in recombination rates. Thus, the recombination rates from the particle simulation data (filled circles) are lower than their initial states (open circles), as shown in Figure \ref{fig:simulation_li}(e).

Figure \ref{fig:simulation_li}(f) displays the ion fractions in ionization equilibrium of the chosen Fe ions. 
The dominant Fe population resulting from accelerated electron distributions is significantly lower than its initial Maxwellian assumption at $T=2 \times 10^7$\ K.
This discrepancy can lead to a substantial overestimation of the plasma temperature deduced from the formation temperature of Fe lines under the assumption of a Maxwellian distribution.
For example, for the accelerated electrons, the population of Fe XXIII is 0.0174, which corresponds to a temperature of $3.98 \times 10^7$\ K under the assumption of a Maxwellian distribution. 
Similarly, the other Fe ions also exhibit higher apparent temperatures of $3.89 \times 10^7$, $3.89 \times 10^7$, and $3.80 \times 10^7$\ K for Fe XXI, Fe XVIII, and Fe XVI, respectively.

\begin{figure}
\centering
\includegraphics[width=1.0\textwidth]{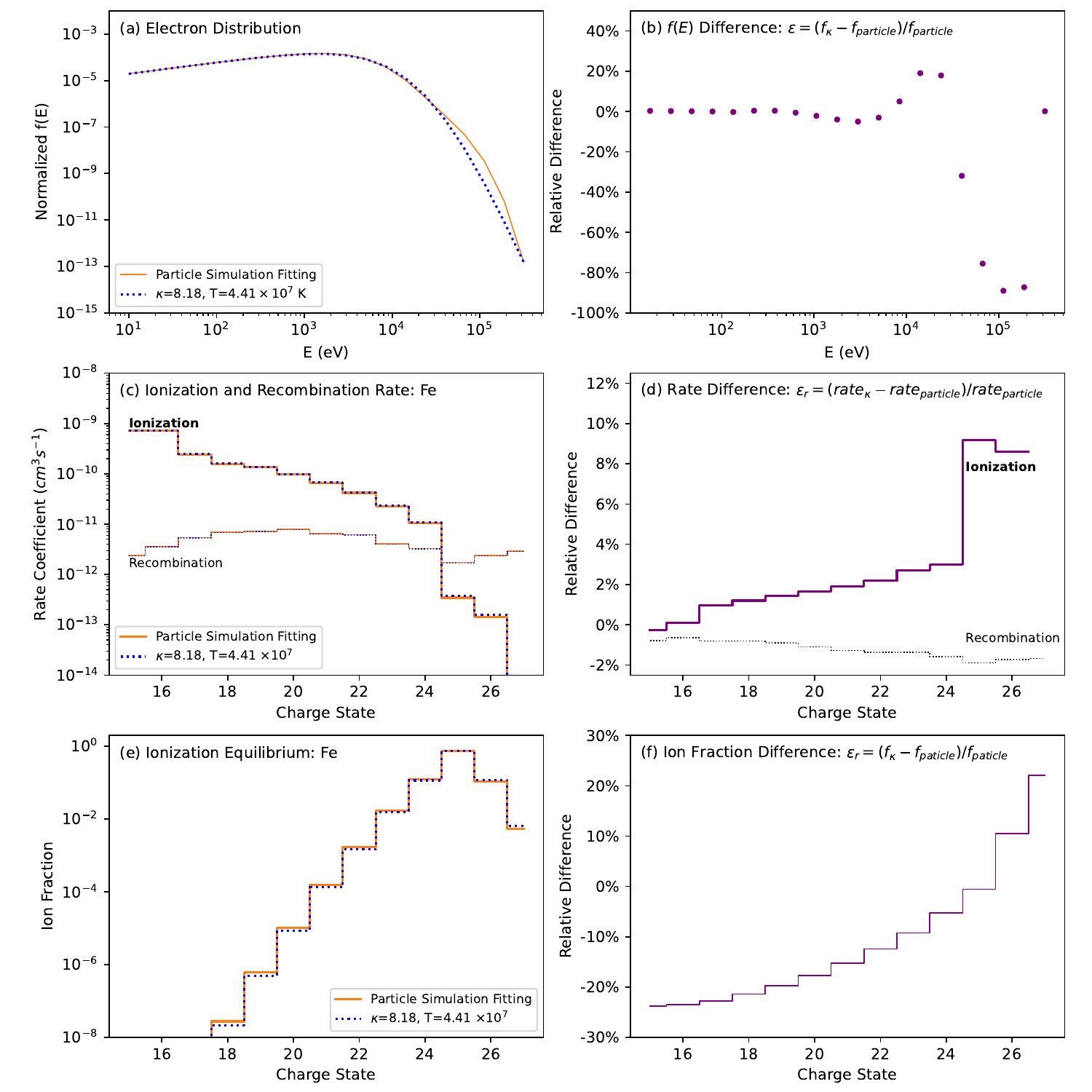}
\caption{Comparison between particle simulation fitting results and the Kappa-distribution approximation.
    (a) Electron distribution profiles obtained from particle simulations (orange solid line) and the best matched Kappa distribution with $\kappa=8.18$ and $ T=4.41 \times 10^7$\ K (blue dotted line);
    (b) Relative difference between the two electron distribution profiles shown in panel (a);
    (c-d) Ionization and recombination rate coefficients for iron in both cases, and their relative differences;
    (e-f) Iron ion fractions in ionization equilibrium states for both cases, and their relative differences.
\label{fig:simulation_li_b}}
\end{figure}

We then examine how deviations appear in ionization analysis when the standard Kappa distribution approximation is used to describe electron profiles in particle simulations.
We first find the best matched $\kappa$ parameters by minimizing the relative difference between any chosen $\kappa$ profile and the particle simulation according to the following formula: 
\begin{equation}
    \text{minimize }  \|\text{mean}(\frac{f_{\kappa}(E) - f_{particle}(E)}{f_{particle}(E)})\|.    
\end{equation}
As shown in Figure \ref{fig:simulation_li_b}(a-b), a standard Kappa distribution with $\kappa=8.18$, and $T=4.41 \times 10^7$\ K is closest to the accelerated electron distribution profiles described above.

The ionization and recombination rates for this chosen Kappa distribution are calculated, and the results are compared with those from the particle simulation fitting, as shown in Figure \ref{fig:simulation_li_b} (c-d).
The overall feature is that both the ionization and recombination rates based on the Kappa approximation basically match the direct particle simulation fitting results. 
The corresponding ion fractions in ionization equilibrium are close to each other, as shown in Figure \ref{fig:simulation_li_b}(e).
Nevertheless, for specific ions such as Fe XXIII and Fe XXVI, the relative differences between the two cases can reach up to $\sim 10\%$ in ionization rate.
Consequently, the ion fractions of these ions in ionization equilibrium show noticeable deviations of up to $\sim 20\%$, as shown in Figure \ref{fig:simulation_li_b}(f).
In summary, the standard Kappa distribution approximation provides a reasonable estimate of ionization and recombination rates as well as ion fractions of these electron distributions. However, the direct Maxwellian decomposition fitting of electron profiles yields even greater accuracy, which is necessary for high-precision ionization simulations.

It is important to note that the electron profile obtained from the above particle simulation is for one time frame during the system's evolution, and for a global electron distribution around the primary magnetic reconnection X-points.
To capture more variation in the accelerated electron distribution, high resolution MHD-particle simulations with a larger number of particles are needed in future work. 
In addition, since our current particle model mainly focuses on exploring nonthermal electron acceleration and transport mechanisms at high energy tails, the detailed particle distribution behaviors in lower energy core are not well addressed yet and require more accurate and self-consistent modeling efforts.
For instance, the injection mechanism of electrons has significant impacts on the low energy distributions, especially in the complex environment around magnetic reconnection sites. 
Therefore, the decrease in recombination rates and the resulting shift of ionization equilibrium states require further exploration in future work.

\subsection{Ionization and Recombination Rates in Solar Wind Modeling with Truncated Electron Distribution Function}
\begin{figure}
\centering
\includegraphics[width=0.8\textwidth]{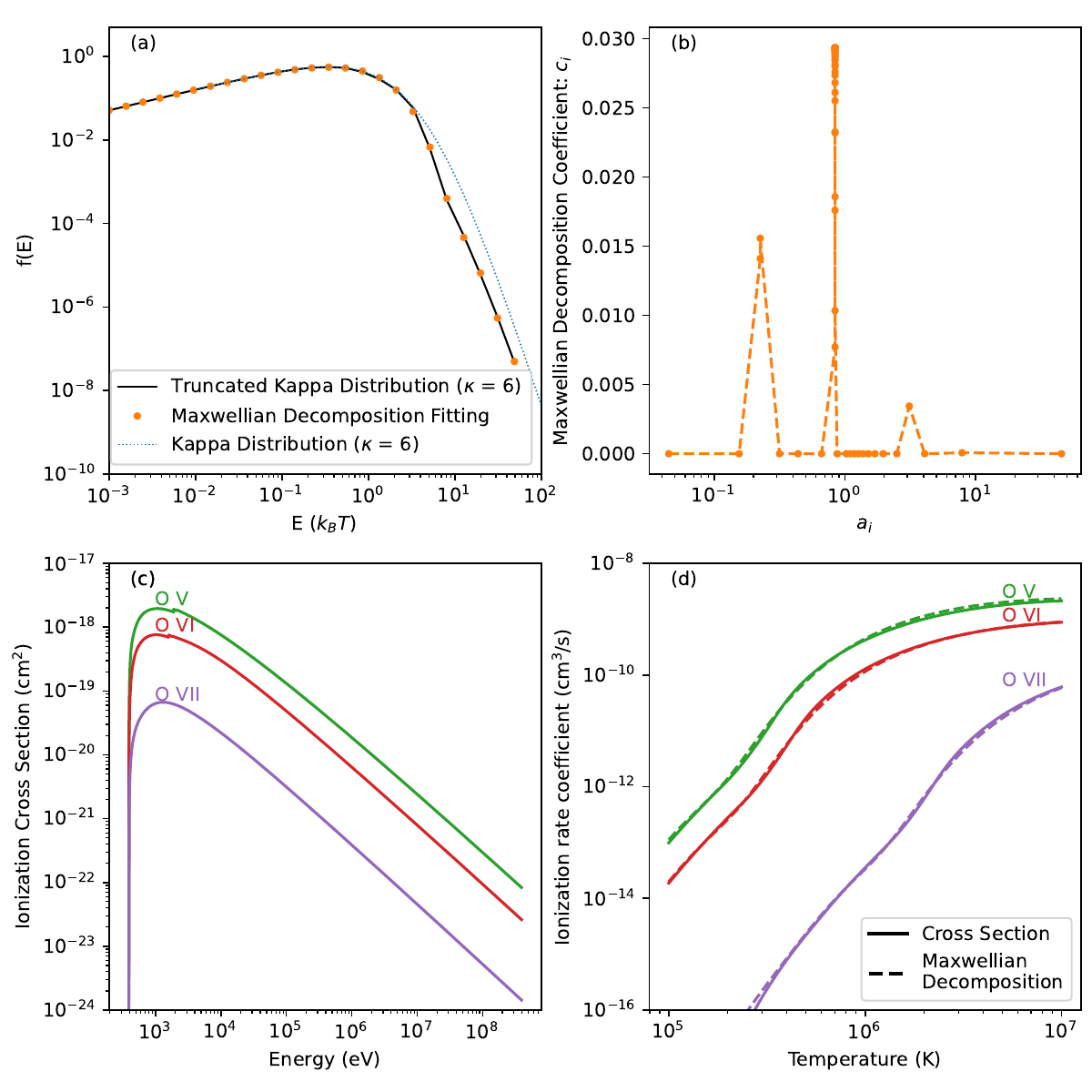}
\caption{The Maxwellian decomposition fitting results for a truncated Kappa distribution ($\kappa=6$) presented in the exospheric model of the solar wind. 
    (a) Electron distribution profiles of the Kappa distribution (cyan dotted line) and truncated case (black solid line). The fitting result is shown as orange dots; 
    (b) The corresponding Maxwellian component coefficients ($c_i, a_i$) used in this fitting; 
    (c) The ionization cross section profiles for the chosen Oxygen ions (O V, O VI, and O VII) are obtained from the CHIANTI database and used to calculate ionization rates; 
    (d) Comparison of ionization rates obtained using two methods: direct integration utilizing ionization cross section data (solid lines) and the Maxwellian decomposition approach (dashed lines).
\label{fig:simulation_slowwind_1}}
\end{figure}

In this section, we apply the Maxwellian decomposition method to a specific electron distribution with truncation at higher energies.
The velocity distribution function in the solar wind is often observed to be anisotropic. In exospheric models \citep[e.g.,][]{Pierrard1996JGR...101.7923P, Pierrard2023_plasma6030036}, anisotropic electron distributions logically exist above the exobase, where collisions can be neglected, and outward-going particles above a certain kinetic energy can escape from the atmosphere. 
As the high energy particles escape and do not return, the velocity distribution function is expected to be truncated, reflecting the absence of high-velocity particles moving toward the solar surface. Above the exobase, there is no electron flux towards the Sun with velocities exceeding the liberation speed \citep{Pierrard2023_plasma6030036}. 
In recent studies, this exospheric model has been applied in the modeling of solar wind acceleration and ion charge state evolutions \citep{Lomazzi2025}.
Figure \ref{fig:simulation_slowwind_1}(a) shows an example of such a truncated electron distribution for $\kappa=6$ at a chosen height ($r=5R_{sun}$). It is noticed that the truncation appears as an abrupt drop at certain high energy in the electron population as derived in the theoretical model. However, due to the numerical difficulty of dealing with these abrupt changes, the truncation is represented by the quick decreasing of $f(E)$ at $E\sim 3k_BT$ in this work as shown by the black line in Figure \ref{fig:simulation_slowwind_1}(a).
Here, the electron energy profile $f(E)$ is obtained by counting both the perpendicular and parallel velocity distribution on the anisotropic electron velocity distribution function (eVDF) map as illustrated by Figure 5 in \cite{Pierrard2023_plasma6030036}.

We then perform Maxwellian decomposition fitting on this truncated $\kappa=6$ profile by utilizing the Linear Regression method with ANNs as described in Section 2.3, and display the fitting results by orange dots in Figure \ref{fig:simulation_slowwind_1}(a). 
The corresponding fitting coefficients ($c_i, a_i$) are shown in Figure \ref{fig:simulation_slowwind_1}(b).
It is worth noting that a negative coefficient ($c_i$) can cause a negative contribution on rate value according to Equation (6). In general, these negative components are expected to be balanced by the dominant positive Maxwellian components, but they might fail when the ionization rate dramatically changes with temperature in the lower temperature ranges (e.g., $T < \sim 10^6$\ K for oxygen ions). In that case, the positive components of the rates at low temperatures may not counterbalance the negative components at high temperatures, as their magnitudes differ by several orders.
In this solar wind case, therefore, we avoid introducing any manifest negative Maxwellian components (less than $-1.0 \times 10^{-5}$) by properly setting the fitting conditions when performing the Linear Regression model.

We calculate ionization rates by using the above fitting coefficients and compare the result with the alternative approach by direct integration utilizing the ionization cross section. 
Figure \ref{fig:simulation_slowwind_1}(c) shows ionization cross section ($\sigma(E)$) as a function of the electron energy for the selected oxygen ions (O V, O VI, and O VII), obtained from the CHIANTI database \citep{Zanna_2021ApJ...909...38D}.
The ionization rate coefficients are then calculated \citep[e.g.,][]{Hahn2015ApJ...809..178H} as:
\begin{equation}
    IonizRate=\int \sqrt{\frac{E}{\mu}} \sigma(E) f(E) dE.
    \label{eq:rate_crosssection}
\end{equation}
Here, $E$ is the energy of the electrons, $\mu$ denotes the reduced mass associated with collisions, which can be approximated by using the electron mass in this work, and $f(E)$ is the truncated electron distribution as shown by the black solid line in Figure \ref{fig:simulation_slowwind_1}(a).
The results of the ionization rate coefficients are shown in Figure \ref{fig:simulation_slowwind_1}(d), where the solid lines denote results by using Equation (\ref{eq:rate_crosssection}) and the dashed lines are results via the Maxwellian decomposition method (Equation 6).
The two results match each other very well, with the maximum relative difference being less than a few percent, even for those small values of the O VII ionization rates.
In general, the calculated rate coefficients via the above two approaches should match each other with very high accuracy, as discussed by \cite{Hahn2015ApJ...809..178H} for the standard Kappa distribution.
Because we chose the Maxwellian decomposition coefficients ($c_i, a_i$) without manifest negative components in Figure \ref{fig:simulation_slowwind_1}, the fitted profile may deviate more obviously and cause additional deviations.
However, the comparison in Figure \ref{fig:simulation_slowwind_1}(d) shows that our Maxwellian decomposition fitting correctly reveals the ionization rate variation in truncated Kappa cases with sufficient accuracy based on the current atomic database.

\begin{figure}
\centering
\includegraphics[width=1.0\textwidth]{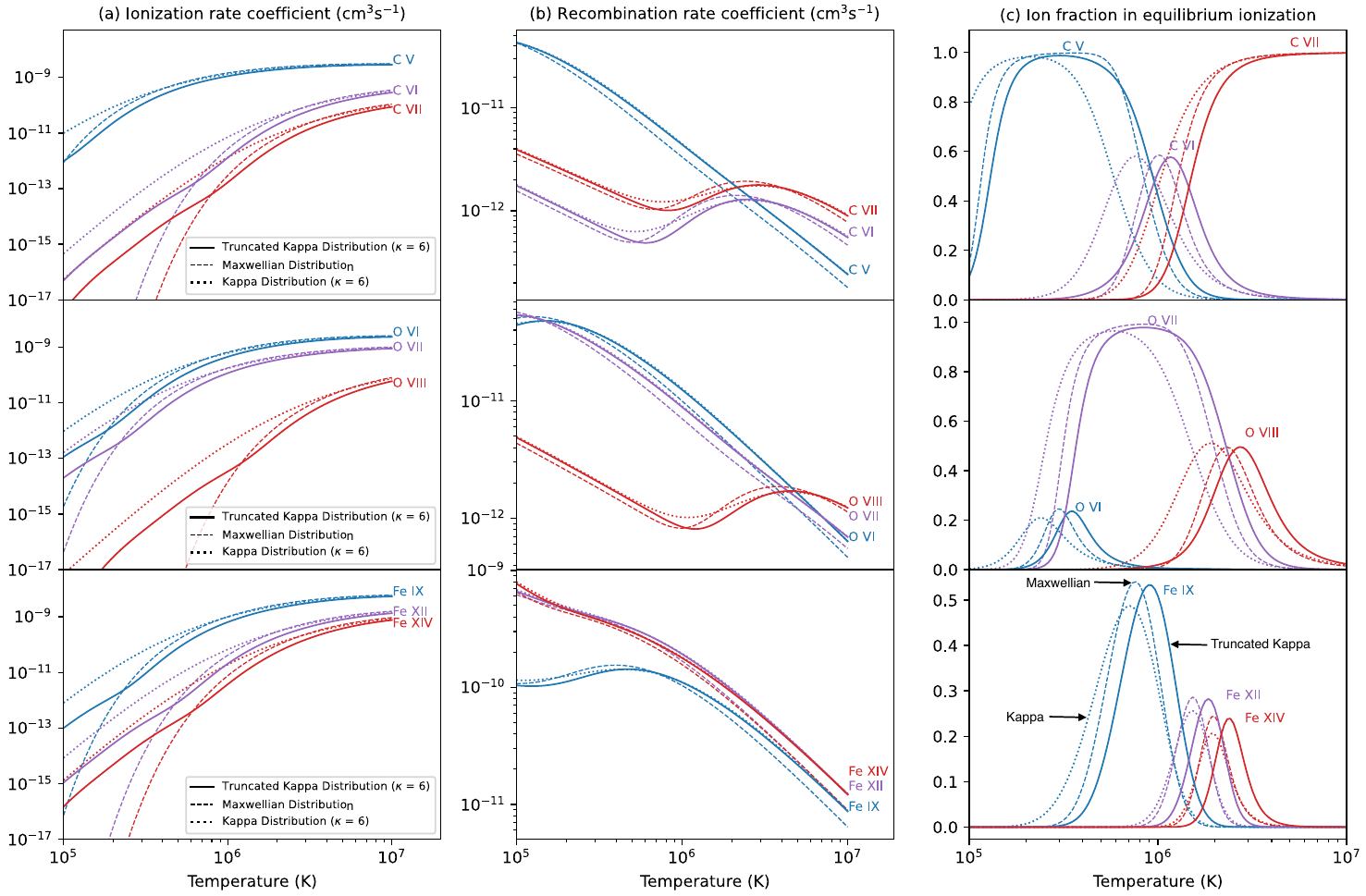}
\caption{Ionization and recombination rates in truncated $\kappa = 6$ distribution in the slow solar wind as shown by solid lines. For comparison, the rates for Maxwellian and standard Kappa distribution cases are shown by dashed and dotted lines, respectively. The different colors indicate the chosen C, O, and Fe ions. The left panels show ionization rate coefficients; the center panels are for recombination rate coefficients, and the right panels show the ion fractions in equilibrium ionization states computed according to these ionization and recombination rates, respectively.
\label{fig:simulation_slowwind_2}}
\end{figure}

Using the above Maxwellian decomposition fitting results and Equation (6), we obtained the ionization rates and recombination rates of the typical ions in the solar wind, and compared the results with both the standard $\kappa$ and Maxwellian electron distributions.
Figure \ref{fig:simulation_slowwind_2}(a) and (b) show the ionization and recombination rate coefficients.
The overall feature is that the Kappa distribution appears larger ionization rates compared with the Maxwellian distribution, while the truncated Kappa distribution appears to have smaller ionization rates than the standard Kappa case, and even lower rates than the Maxwellian assumption in some temperature ranges (e.g., $T > 8 \times 10^5$\ K for C VII).
For recombination rates of most ions, the deviations of both the standard Kappa and truncated Kappa distribution cases from the Maxwellian distribution are generally small, on the order of $\sim$0.1 in relative difference. 
An exception exists in ions such as C V, C VI, and O VII, which show notable differences around $T \sim 10^6$\ K compared to both the Maxwellian and $\kappa=6$ cases.

The ion fractions in ionization equilibrium states are calculated accordingly and shown in Figure \ref{fig:simulation_slowwind_2}(c).
In a standard Kappa case, the ion fraction is mainly characterized by an under-ionized feature, which can be seen from the shifting of the ion population profiles (represented by dotted lines) towards a lower-temperature charge state when compared to the Maxwellian assumption (displayed by dashed lines).
In contrast, in the truncated Kappa case, an over-ionized feature is observed, as indicated by the ion population profile (solid lines) shifting toward higher temperature.
For example, the peak of the population of truncated $\kappa=6$ for O VI clearly shifts to the higher temperature at T $\sim$354000\ K compared to these in both the Maxwellian assumption (at $\sim$290000\ K) and standard $\kappa=6$ case (at $\sim$240000\ K).
Furthermore, for ions such as Fe IX, XII, and XIV, the dominant temperatures are also systematically higher than those predicted by the Maxwellian assumption by a factor of $\sim$1.2.
These departures of truncated ion population profiles from Maxwellian distribution could significantly impact plasma diagnostics in the solar wind, and potentially lead to an over-estimation of plasma temperature.
Considering that the truncation varies with different heights from the solar center and also changes with different $\kappa$ values (as well as the Maxwellian distribution), this could help better understand the evolution of the ionic charge states in the solar wind \citep{Ko2023AGUFMSH51E2653K}. Further exploration of the impact on the ionization and recombination processes by these truncated distributions is required in future studies.

\section{Summary}
In this paper, we describe the Maxwellian decomposition method used to calculate ionization and recombination rates for various non-Maxwellian electron distributions. In particular, we develop two models to fit the standard $\kappa$ distribution and arbitrary non-Maxwellian distributions: (I) the classic iterative model and (II) a linear regression model enhanced by ANN.
Our results show that both models perform well, with accurate fitting results comparable to those from previous tables, showing only minor differences. 
Additionally, the new methods support more flexible application environments and can be easily applied to updated atomic databases.
Thus, these new approaches are reliable and efficient for conducting further ionization analysis.

\begin{itemize}

    \item {We obtained the Maxwellian decomposition coefficients $c_i$ and $a_i$ for the standard $\kappa$ distribution and calculated the corresponding ionization and recombination rates by summing over the Maxwellian components.
    By comparing different fitting errors, particularly at the high energy end, we emphasize that high-accuracy fitting is necessary to obtain precise ionization and recombination rates. For a typical low $\kappa$ case,  $\kappa=2$, all Maxwellian components with $c_i > \sim 10^{-5}$ contribute significantly; 
    For a high $\kappa$ case, $\kappa=10$, the ionization rates are still impacted by minor $c_i$ components, but ion populations in ionization equilibrium are less affected. }

    \item {We calculated the ion populations of carbon, oxygen, and iron in ionization equilibrium and compared our results with others, including \cite{Hahn2015ApJ...809..178H} tables and KAPPA packages. Our results closely match those, with a relative difference of only a few percent for the dominant carbon and oxygen ions. 
    However, the iron population shows some differences compared to the KAPPA package when the population drops to $\sim 10^{-3}$, but the peak fractions are close to each other.}
    
    \item {We apply our fitting methods to two particular non-Maxwellian electron distribution scenarios: (I)including segmented high energy tail in magnetic reconnection sites in solar flares, (II)including high energy truncation in the exospheric model of the solar wind.}
    
    \item {Based on the particle simulation results, we obtain the electron distribution and find that the high energy tails generally evolve with time and cannot be well described by using a single standard $\kappa$ function.
    Compared to its initial electron Maxwellian distribution, the ionization rate around the reconnection sites is significantly higher due to the contribution of high-energy tails, while the recombination rates may decrease due to the electron distribution changes in low-energy ranges. This results in a lower ion fraction of those high-temperature Fe ions (e.g., Fe XVIII to Fe XXIII) compared with the Maxwellian distribution assumption, potentially leading to an overestimation of plasma temperature when analyzing the Fe emission lines.}

    \item {In an exospheric model of the solar wind with high energy truncation, the ionization rates are significantly lower than those in the standard $\kappa$ distribution, while the recombination rates are close to each other. 
    Thus, due to the reduced ionization rates, the typical O VI and O VII ion populations with truncated electron distribution are more than an order of magnitude lower than those in the untruncated Kappa case}.

\end{itemize}

Performing the direct Maxwellian decomposition on non-Maxwellian distributions other than the standard $\kappa$ distribution provides a useful tool in various applications.
In these environments with energetic particles, fitting ionization and recombination rates is much more complex as the electron distribution function is in time-dependent evolution.
Thus, the quick Maxwellian decomposition fitting is required to address the actual high-energy tail departure from the power-law distribution.
On the other hand, the low energy core also notably contributes to the recombination rates. To our knowledge, there are no comprehensive studies of how the electron density changes in the low-energy Maxwellian core during electron acceleration, which could significantly affect the estimation of recombination rates.
In practice, the electron injection mechanism during particle acceleration processes is thought to be an important factor, which requires more investigation.

We noticed that our model shows some differences in ion population in ionization equilibrium, compared with results from the KAPPA package.
The discrepancy is probably caused by the inaccuracy of recombination rates in both methods. The uncertainty is from reverse-engineering-generated dielectronic recombination rate coefficients in the KAPPA packages, and extrapolation of the rates in the Maxwellian decomposition methods, which leads to inaccuracy in the rate coefficients.
However, the largest relative differences (up to $\sim 50$\%) only appear in the Fe ions with insignificant populations at the level of $10^{-3}$.

The two fitting models utilized in this work have their advantages and shortages.
Using a classic iterative model, one can efficiently fit electron distributions of the standard $\kappa$ distribution with $\kappa$ values ranging from 1.7 to 100 with the extremely high accuracy of the double precision of float type.
The fitting framework can also be extended to other fitting processes beyond Maxwellian decomposition. 
In fact, applying other linear compositions is straightforward: simply replace the core function from a Maxwellian function to another core, such as the $\kappa$ function.
On the other hand, the specified linear regression model optimized by the
artificial neural network (ANN) is more flexible in various applications. 
One of the advantages of performing a neural network model is that the weights of each Maxwellian component can be easily constrained. For instance, the unphysical components with $c_i < 0$ are necessary to fit some particular distribution profiles with energy truncations. 
However, the negative $c_i$ components might cause a negative contribution to ionization and recombination rate coefficients.
In such cases, the constraints on the range of $c_i$ are helpful during the training process by using the neural network model.
In summary, the direct training processes might require more computing time (e.g., tens of seconds) than the classic iterative model and also require more effort to find proper optimizers.
Thus, we suggest using the classic iterative model as the primary fitting model, as it supports the highest fit accuracy and costs fewer computer resources.
Meanwhile, the neural network model can be used to assist in finding proper parameters in complex tasks. The neural network model also benefits from the quick development of AI technology and might support more convenient and efficient fitting in the near future.

The truncated electron distribution might have significant impacts on the Fe charge states modeling in solar winds. 
As shown in Figure \ref{fig:simulation_slowwind_2}, the suprathermal electron tail, such as $\kappa$-distribution, can efficiently increase the ionization rates and lead the equilibrium ionization states to shift to the low-temperature side. 
Thus, the Fe charge states with non-thermal electrons in $\kappa$ distribution can be expected to show lower charge states than the Maxwellian assumption.
However, there is no clear evidence about the obvious drops of average Fe charge states from in-situ measurements in solar wind studies associated with suprathermal strahl \citep[e.g.,][]{Rivera2021ApJ...921...93R}. 
In contrast, observed Fe charge states might be even higher than the modeling prediction in the Maxwellian distribution assumption.  \cite{Lionello2019SoPh..294...13L} reported ion charge states model combined in the wave-turbulence-driven model of the solar wind. They conducted an analysis of time-dependent ionization for the fast solar wind and compared model predicted ion populations with measurements from the Ulysses spacecraft. They found that the iron charge states are generally lower than expected in models with corrected ion outflow speed, while the other charge states of carbon and oxygen matched well with the Ulysses measurements.
Considering that Fe ions could freeze-in at higher altitudes than carbon and oxygen, the truncated electron distribution can further affect the Fe evolution and might cause higher charge states than these frozen-in carbon and oxygen. In addition, the evolution of average Fe charge states and other typical solar wind ion ratios, such as O$^{+7}$/O$^{+6}$, also has been noticed in other models \citep[e.g.,][]{Szente2022ApJ...926...35S}.
Thus, further time-dependent ionization modeling with truncated high-energy tail could help better understand the evolution of Fe charge state in the solar wind.

As discussed above, our results thus show that properly accounting for non-Maxwellian conditions in the solar plasma is crucial for the interpretation plasma diagnostics from both current (e.g. EIS, IRIS, Solar Orbiter, Parker Solar Probe) and future (e.g. MUSE, EUVST) solar missions.
\begin{acknowledgments}
This work is supported by NASA grants 80NSSC21K2044 to the Smithsonian Astrophysical Observatory. C. Shen acknowledges the support from NASA through Grant 80NSSC25K7707, 80NSSC20K1318, and NSF AST 2108438. X. Li acknowledges the support from NASA through Grant 80NSSC21K1313, National Science Foundation Grant No. AST-2107745, and Smithsonian Astrophysical Observatory through subcontract No. SV1-21012. Y.-K. Ko acknowledges support by the Office of Naval Research. V. Polito acknowledges support from NASA grants 80NSSC21K2044. V. Pierrard acknowledges the project 21GRD02 BIOSPHERE from the European Partnership on Metrology, co-financed by the European Union’s Horizon Europe Research and Innovation Programmer and by the participating states. Y.-K. Ko and V. Pierrard acknowledge insightful discussions within the International Team “Heliospheric Energy Budget: From Kinetic Scales to Global Solar Wind Dynamics” at the International Space Science Institute (ISSI) in Bern led by M. E. Innocenti and A. Tenerani.

\end{acknowledgments}

%



\software{astropy \citep{2013A&A...558A..33A,2018AJ....156..123A},  
          CHIANTI \citep{Dere_2019ApJS..241...22D}, 
          Kappa Package \citep{Dzifcakova2021ApJS..257...62D},
          Scikit-learn \citep{scikit-learn}
          }



\appendix
The Maxwellian decomponent fitting code used in this work has been made available to the public and can be freely accessed in an open-source program\footnote{ https://github.com/ionizationcalc/Non-MaxwellianDistribution.}, which also includes the tabled ionization and recombination rates.

\begin{figure}
\centering
\includegraphics[width=0.8\textwidth]{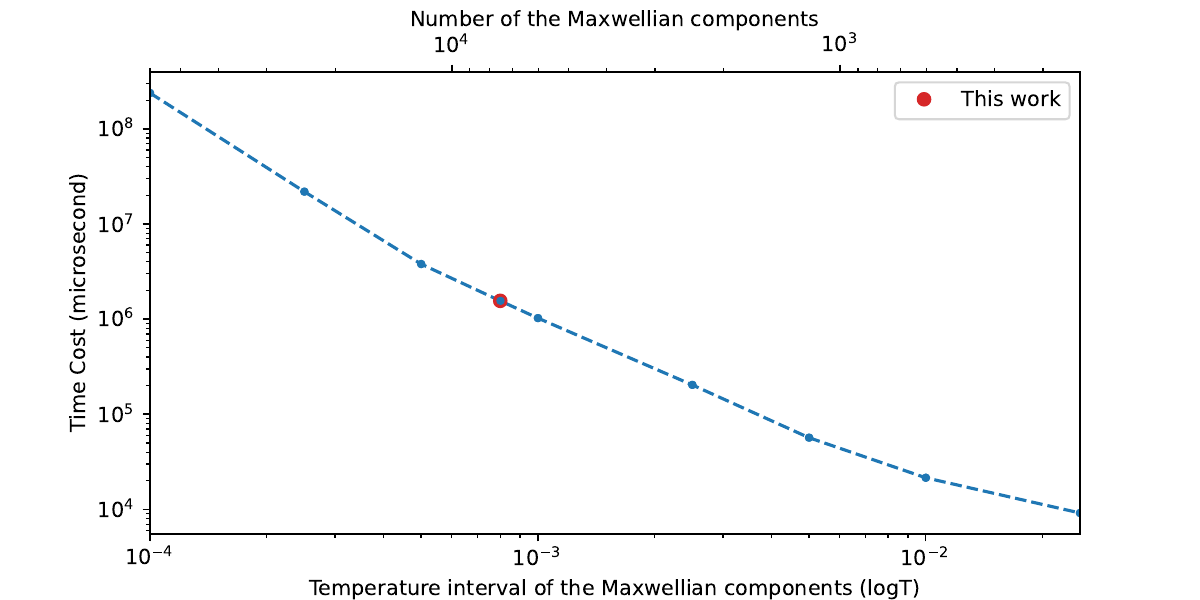}
\caption{The time cost for performing linear fitting model using single CPU core of Intel(R) Xeon(R) Silver 4214 running at 2.2GHz. The lower horizontal axis shows the interval of temperature grids in a logarithmic scale, and the upper horizontal axis is the corresponding sampling number of all Maxwellian components. The red cycle indicates the sampling number we utilized to create Figure \ref{fig:linearfit} in this work.
\label{fig:linearfit_timecost}}
\end{figure}

As shown in Figure \ref{fig:linearfit_timecost}, we examined the computational time-costs for the classic iterative method described in Section 2.2. The test was performed on a commonly used X86 platform equipped with an Intel(R) Xeon(R) Silver 4214 CPU at 2.2GHz. The results show an achievable calculation speed of a few seconds with $10^4$ Maxwellian components. However, as the number of Maxwellian components approaches $\sim 10^5$, the time cost increases significantly.

\begin{figure}[ht!]
\centering
\includegraphics[width=0.5\textwidth]{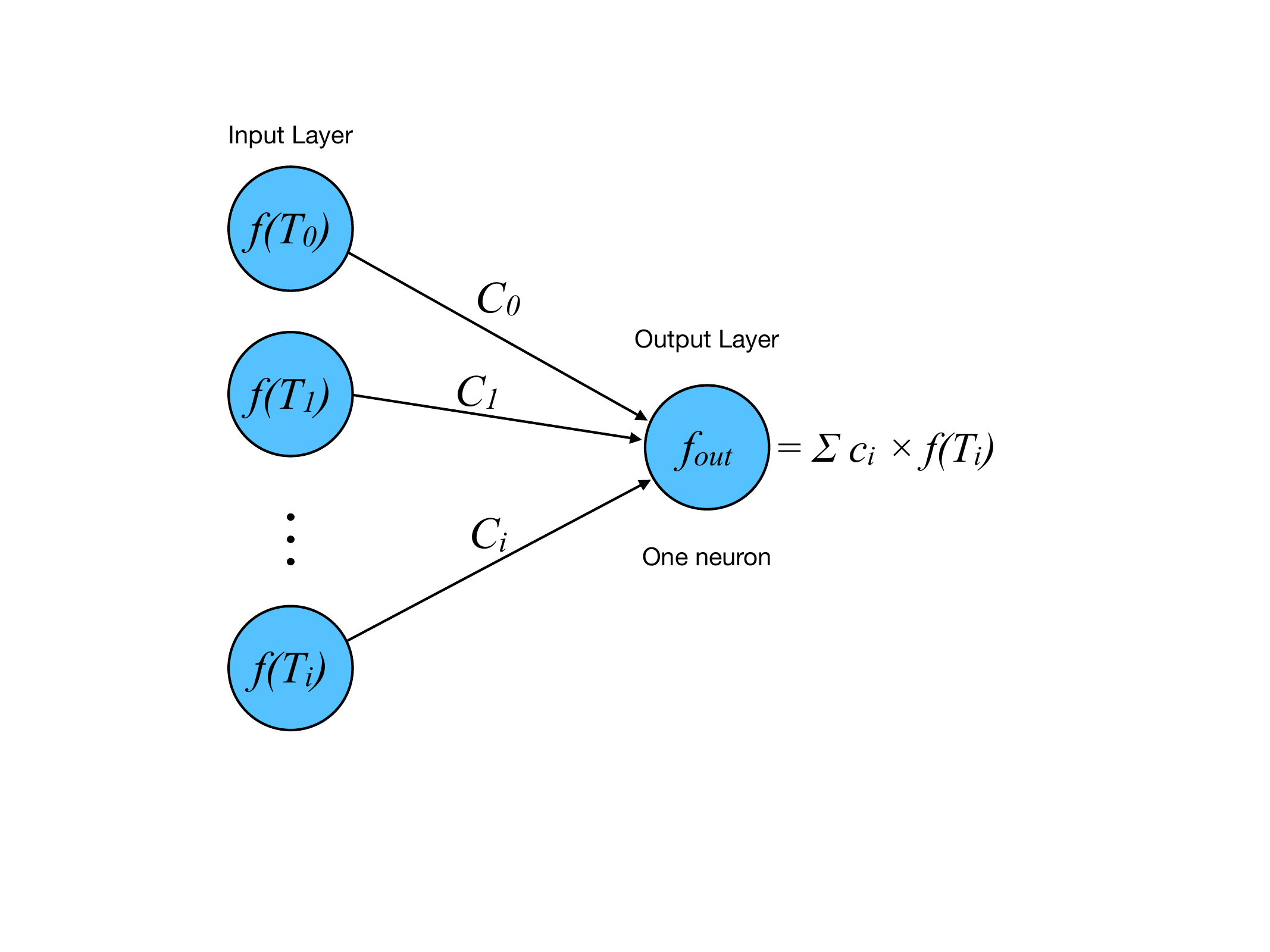}
\caption{The single-layer neural network employed in this work. The input layer is represented by multiple Maxwellian distributions ($f(T_i)$) corresponding to different temperatures $T_i$. The output layer shows the sum of these Maxwellian distributions with different weights ($c_i$), and no hidden layers in this model.
\label{fig:neuron}}
\end{figure}

Figure \ref{fig:neuron} illustrates the connectivity pattern of the one neuron in the linear regression model with ANNs described in Section 2.3. We noticed that our linear regression model is different from the general ANN model. The training, in our case, means actively changing the weights of linear components (e.g., $c_i$). The training aims to find the best linear components that match the inputs and return these weights. 
On the other hand, other commonly referenced ANN models, such as deep learning models, offer a different approach. These ANN models generally collect characteristic features from a large number of samples and finds the proper hidden parameters in multiple neural network layers to describe the relationship between the input features (e.g., electron distribution profiles) and the output values (e.g., Maxwellian components coefficients).
The training of these deep-learning models is used to find all these hidden parameters, where the number of hidden parameters largely depends on the complexity of the applications. As a result, training a deep-learning model typically requires more computational resources than a linear regression model.
However, once the training is completed, using the trained deep-learning model to estimate the best Maxwellian component coefficients is much faster. 
Therefore, further studies could benefit the development of such deep-learning models with sufficient samples.
In fact, the two fitting methods reported in this paper can also assist in setting up these samples for deep-learning models. For instance, one can set up the distributions with various $\kappa$ values and perform linear fitting, as discussed in Sections 2.2 and 2.3, to obtain both the Maxwellian component coefficients and electron profiles as training datasets for deep-learning models. Alternatively, a sufficient number of electron distributions can be collected directly from particle simulations, as mentioned in Section 3.2, to prepare the training dataset accordingly.
By providing sufficient samples covering various electron distributions, we expect that the deep-learning model will give a proper prediction of the Maxwellian components for arbitrary electron distribution profiles, ensuring acceptable fitting accuracy in future studies.

\bibliography{ref}
\bibliographystyle{aasjournal}



\end{document}